\documentclass[review]{elsarticle}

\usepackage{hyperref}
\usepackage{lineno}
\usepackage{multirow}
\usepackage{multicol}
\usepackage{gensymb}
\usepackage{amssymb}
\usepackage{subcaption}
\usepackage{tablefootnote}
\usepackage{caption}
\usepackage{subcaption}
\usepackage{wrapfig}

\graphicspath{ {pictures/} }

\newcommand{\mum}{~$\mathrm{\mu m}$}

\newcommand{\simile}{$\sim$}

\usepackage{xcolor}

\journal{NIM Section A}

\begin{document}
\begin{frontmatter}

\title{First experimental results of the spatial
resolution of RSD pad arrays read out with a 16-ch board}

\author[address1,address3]{F. Siviero\corref{mycorrespondingauthor}}
\cortext[mycorrespondingauthor]{Corresponding author}
\ead{federico.siviero@edu.unito.it}
\author[address6]{F. Giobergia}
\author[address3,address1]{L. Menzio}
\author[address5]{F. Miserocchi}
\author[address1]{M. Tornago}
\author[address4]{R. Arcidiacono}
\author[address3]{N. Cartiglia}
\author[address1]{M. Costa}
\author[address4]{M. Ferrero}
\author[address1]{G. Gioachin}
\author[address3]{M. Mandurrino}
\author[address3]{V. Sola}

\address[address1]{Università degli Studi di Torino, Torino, Italy}
\address[address3]{INFN, Torino, Italy}
\address[address4]{Università del Piemonte Orientale, Novara, Italy}
\address[address5]{D-ITET, ETH, Zürich, Switzerland}
\address[address6]{Politecnico di Torino, Torino, Italy}

\begin{abstract}

Resistive Silicon Detectors (RSD, also known as AC-LGAD) are innovative silicon sensors, based on the LGAD technology, characterized by a continuous gain layer that spreads across the whole sensor active area. RSDs are very promising tracking detectors, thanks to the combination of the built-in signal sharing with the internal charge multiplication, which allows large signals to be seen over multiple read-out channels. This work presents the first experimental results obtained from a 3$\times$4 array with 200~\mum~pitch, coming from the RSD2 production manufactured by FBK, read out with a 16-ch digitizer. A machine learning model has been trained, with experimental data taken with a precise TCT laser setup, and then used to predict the laser shot positions, finding a spatial resolution of $\sim$~5.5~\mum. 

\end{abstract}

\begin{keyword}

LGAD, AC-LGAD, Particle tracking detectors (Solid-state detectors)

\end{keyword}

\end{frontmatter}

\tableofcontents


\section{Introduction}
RSDs (Resistive AC-Coupled Silicon Detectors) are $n$-in-$p$ silicon sensors based on the LGAD (Low-Gain Avalanche Diode) technology, featuring an unsegmented gain implant that spreads across the whole sensor active area. The two key characteristics of the RSD design are the built-in signal sharing and the internal charge multiplication, which allow the RSDs to precisely reconstruct the hit positions of ionizing particles. A detailed description of the RSD design and its principles of operation can be found in~\cite{MM2,marta}.

Two RSD productions, RSD1 and RSD2, have been manufactured at Fondazione Bruno Kessler (FBK) between 2019 and 2021~\cite{MM_RSD1,MM_RSD2}; the results in this work are related to sensors belonging to the latest RSD2 batch.

Previous works~\cite{fede_ml,marta} focused on a subset of 3-4 read-out channels to reconstruct the particle hit positions with the RSD, whereas, in this article, the whole sensor (for a total of 12 channels) has been read out, finding similar results and proving, in this way, the consistency of the methodology adopted.

\section{Laboratory measurements}

RSD2 sensors have been studied in the Torino LISS, the Laboratory for Innovative Silicon Sensors, with the Particulars Transient Current Technique (TCT) setup~\cite{TCT}. The setup comprises: (i) a picosecond infrared laser with 1060~nm wavelength and a spot of \simile~8~\mum, and (ii) a moving $x-y$ stage with sub-micron precision, on which the device-under-test (DUT) is mounted. The infrared laser well simulates the passage of a minimum-ionizing-particle (MIP) in the sensor; the movable $x-y$ stage provides the reference position of the laser shot with high accuracy ($\sigma_{laser}$~\simile~2~\mum) and allows for a detailed mapping of the DUT.

RSDs were wire-bonded to a 16-channel read-out board, designed at the Fermi National Accelerator Laboratory (FNAL). All pads were read-out simultaneously with a Caen 16-channel digitizer. A dedicated software was developed to equip the setup with an automatic data acquisition system, which allows remote control of the power supply, the TCT moving stage, and the digitizer. 

\begin{figure}[h]
\begin{center}
\includegraphics[width=9cm, height=4cm]{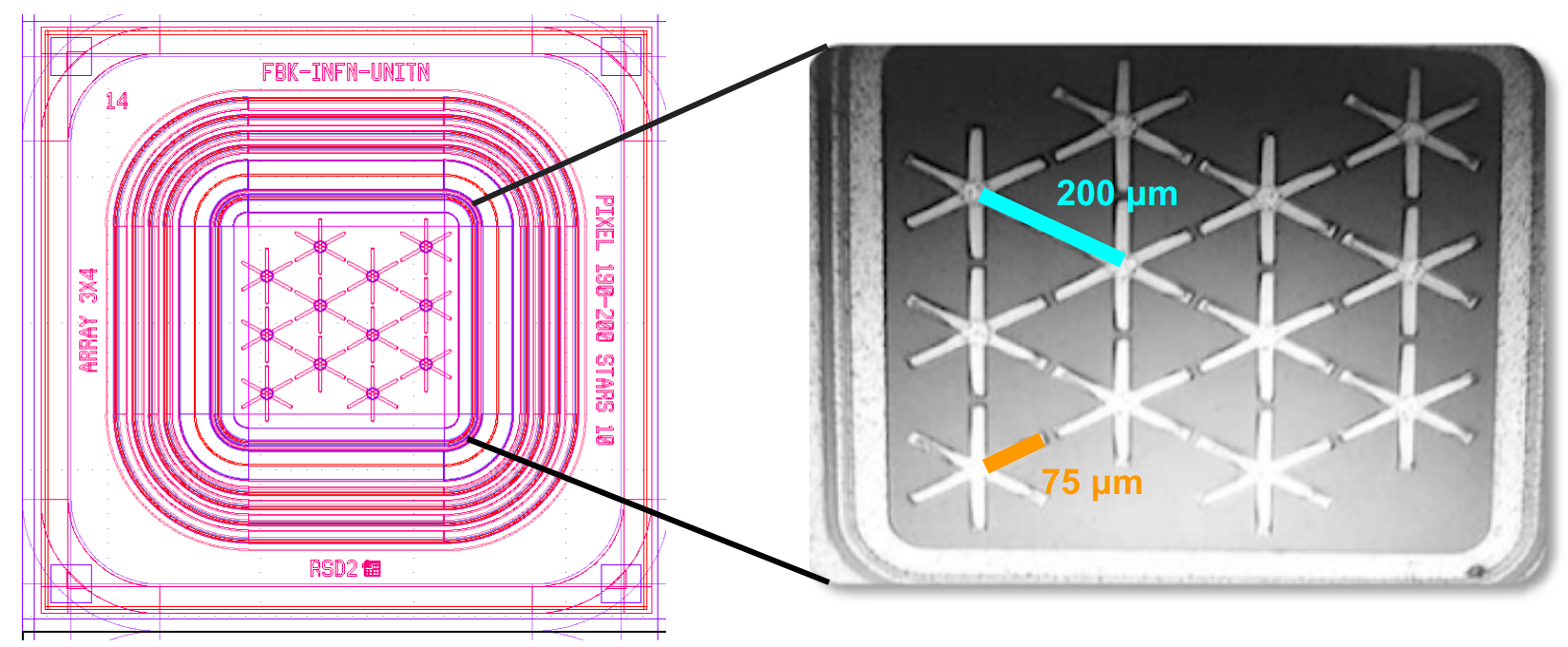}     
\caption{$Left$: Schematic view of the DUT. $Right$: picture of the DUT active area, taken with a microscope.}
\label{fig:sensor_pic}
\end{center}
\end{figure}

The measurements consist of a scan of the sensor active surface with the TCT laser. The scan follows a grid of points spaced by 10~\mum~in both $x$ and $y$, whose position is provided by the TCT $x-y$ moving stage and recorded by the acquisition system; 100 signal waveforms are acquired in each position, for a total of 1.65$\cdot$10$^{6}$ events. 

The DUT in this work, shown in figure~\ref{fig:sensor_pic}, is a 3~$\times$~4 array with an active thickness of 50~\mum and an active area of 750~$\times$~750~\mum$^{2}$, featuring cross-shaped metal read-out pads with 200~\mum~pitch, and 75~\mum-long and 20~\mum-wide arms. The sensor has been biased at 250~V, corresponding to an internal gain of \simile~20.

One of the reasons to produce a sensor with cross-shaped read-out pads is to minimize the area covered by metal: signal sharing is fundamental in RSD to precisely reconstruct the particle hit position, but the sharing does not occur when the particle hits the metal (only the hit pad sees the signal in that case). In addition, with this particular geometry, most of the signal is contained in the 3 read-out pads contouring the hit position, ensuring better performance and higher uniformity: it is known from the first RSD production (RSD1), in fact, that the sharing involving a too large (or non-constant) number of pads worsens the reconstruction~\cite{cern-seminar}.
 
\begin{figure}[h]
\begin{center}
\includegraphics[width=\linewidth]{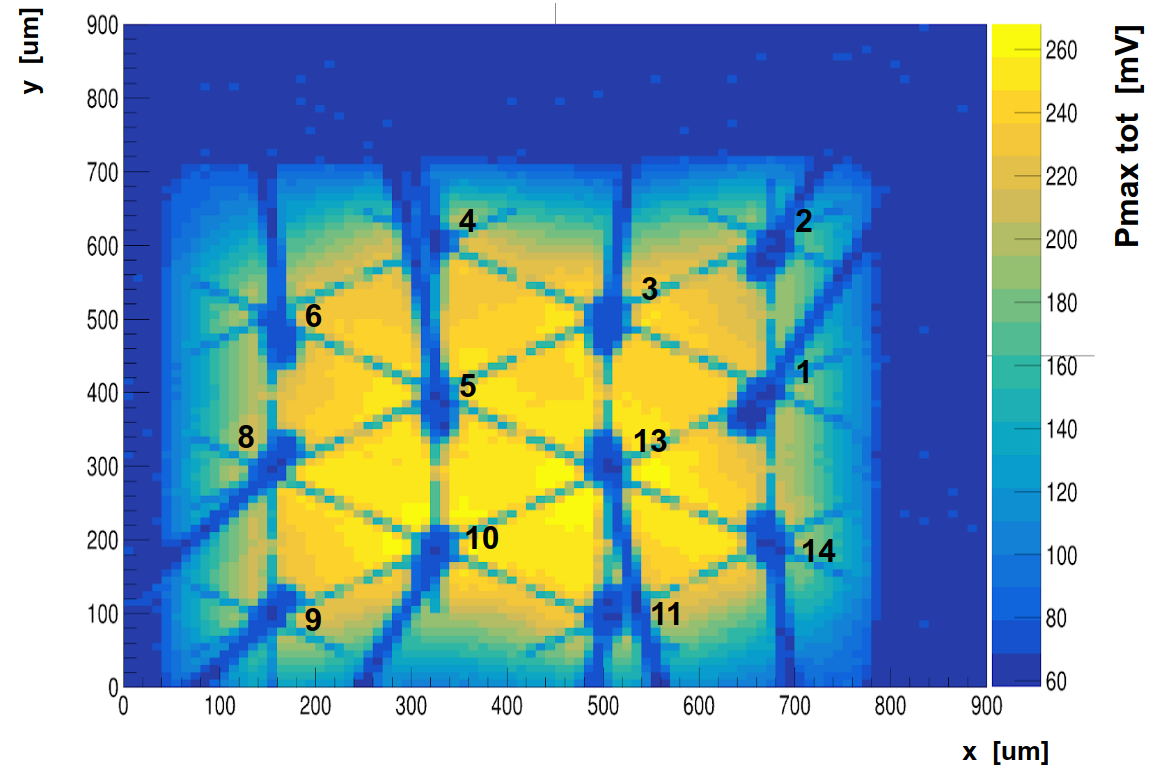}     
\caption{$x-y$ map, obtained with the TCT setup, of the DUT. The colored axis presents the total amplitude, obtained by summing the amplitudes of the signals seen by each read-out pad when the laser shots on a given position.}
\label{fig:tct_map}
\end{center}
\end{figure}

Figure~\ref{fig:tct_map} shows a $x-y$ map, obtained with a TCT scan, of the DUT active area: the colored axis presents the total amplitude ('Pmax tot'), obtained by summing the amplitudes of the signals seen by each pad when the laser shoots on a given position. The pad numbers are also shown.

Although all the positions in the 0-900~\mum~range (both in $x$ and $y$) have been measured, only the positions within the yellowish region in figure~\ref{fig:tct_map} have been considered for the reconstruction process, since the performance of the sensor in the periphery is different given that there the signal sharing pattern is different. 

The blue regions in the map are those where the laser is absorbed, so the signal amplitudes are zero or close to zero. The blue regions correspond to the DUT metal read-out pads and to the wire bonds (those blue strips crossing the active area in various positions). Even though the wire bonds are only 20~\mum~wide, they are clearly identified by the TCT: this highlights the good accuracy of this setup, which is fundamental to effectively measure the properties of RSDs.

\begin{figure}[h]
\begin{center}
\includegraphics[width=0.65\linewidth]{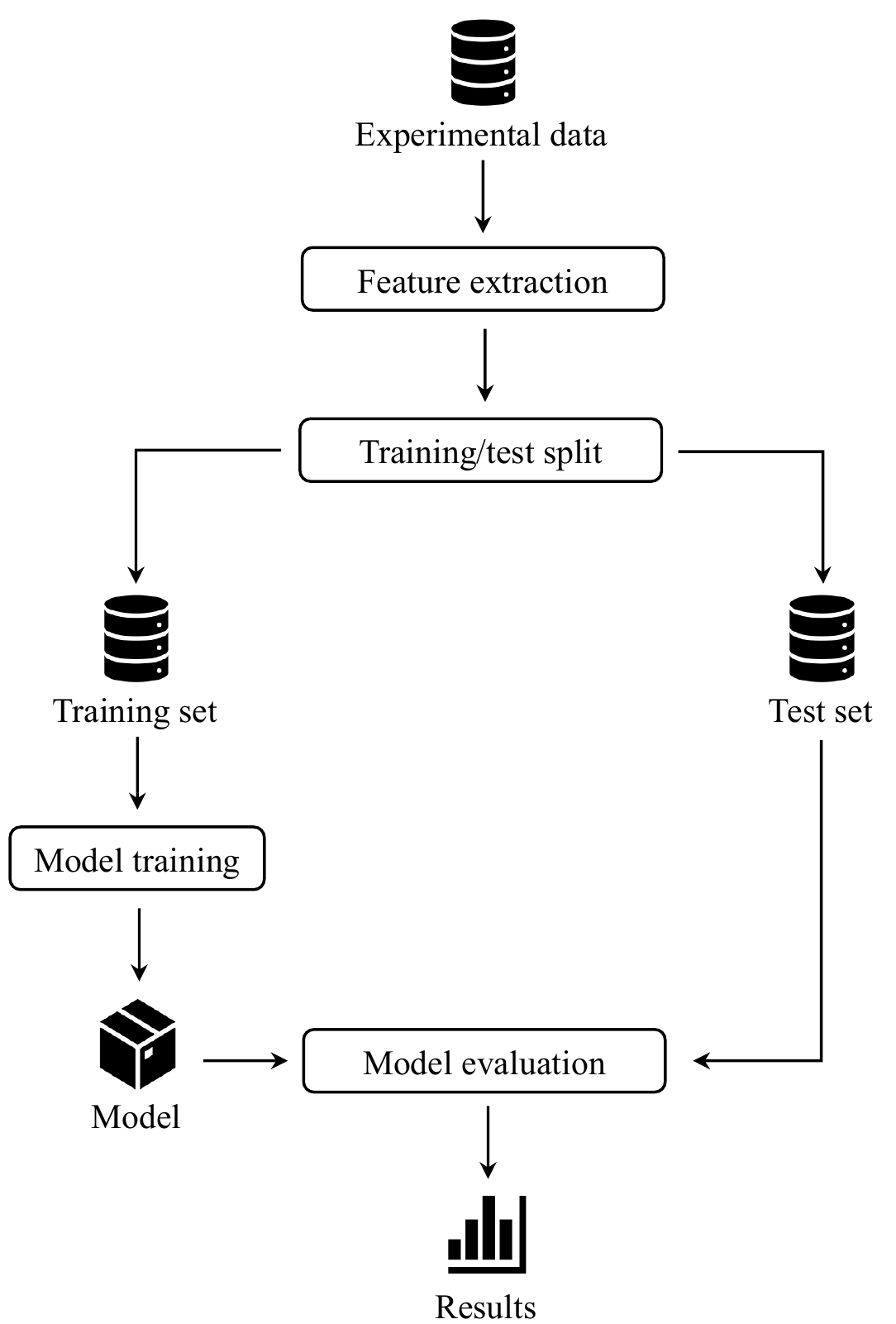}
\caption{Data science pipeline adopted.}
\label{fig:pipeline}
\end{center}
\end{figure}

\section{Position reconstruction and Machine Learning}

The position reconstruction process with RSDs relies on the internal built-in signal sharing: when an ionizing particle hits the sensor, the neighbouring pads see a signal; these signals carry a lot of information, such as the amplitude, the area, the width, the slew rate, which can be used to accurately reconstruct the hit position. This is the RSD "recipe": the combination of multiple analog read-out channels leads to a drastic improvement in the spatial reconstruction.

The reconstruction process is rather straightforward: the signal characteristics are valuable features to be provided as an input, while the $x-y$ coordinates of the hit position are the expected outputs. However, a similar approach is very challenging when relying on analytical laws, which are complex and not easy to derive~\cite{fede_ml}. This naturally calls for the use of machine learning (ML) algorithms: the algorithm can be trained with the signal characteristics as input features, and then, once it is trained, it can be used to predict the $x-y$ coordinates of the hit position, with no need to know complex analytical laws.

The methodology adopted for the reconstruction process can be summarized through a standard data science pipeline: figure~\ref{fig:pipeline} highlights the most important steps taken, as well as the intermediate results produced. In particular, the raw data collected are firstly processed to extract meaningful features for the regression model (the so-called \textit{feature extraction}); then, the dataset is split into two subsets, used for the training and testing of the model, respectively (\textit{training / test split}). During the learning process (\textit{model training}), a regression model is fit using the training subset. After training, the performance of the model is assessed on a disjointed set of data (\textit{model evaluation}). The following paragraphs delve into the details of each of the aforementioned steps of the pipeline. The model evaluation will instead be discussed in section~\ref{sec:conclusions}.


\textit{Feature extraction}: the raw data acquired with the digitizer are analysed before feeding them to the ML algorithm. We used the 12 signal amplitudes recorded by the read-out channels in each event as input features. We discarded all the events where the laser is absorbed, i.e. metal pads and wire bonds. The output of the feature extraction phase is a tabular dataset, with a row for each of the recorded events and a column for each of the 12 features. In other words, each event is described by a point $\mathbf{v} \in \mathbb{R}^{12}$.

\textit{Training / test split}: to guarantee a fair evaluation of the regression model, the available dataset is split into two non-overlapping parts: a training set (80\% of the dataset) and a test set (20\% of the dataset). The training set is used for the training phase of the model, whereas the test set is used to assess the quality of the model on never-before-seen data. 

It is important not to use data that have already been leveraged for the training phase, in order to assess the generalization capabilities of the trained model. This is the reason to have non-overlapping training and test sets. As an additional guarantee of separation between the two sets, we assigned all events recorded for a specific pair of coordinates to one or the other set (i.e. all samples for a given ($x,y$) pair either belong to the training set or the test set). This implies that, during the test phase, the model will need to predict coordinates that have not been used for the training of the model, as would be the case when deployed.

\textit{Model training}: a random forest has been used as a regression model. A random forest is an ensemble model where multiple decision trees (in our case, 100) are trained in parallel. More specifically, each tree is trained on a random sample of the original training set, and, for each split of the tree, only a random subset of columns is used. The predictions of all trees are then averaged to obtain the overall prediction. This has been shown to result in reduced variance (i.e. less prone to overfitting) when compared to decision trees \cite{breiman2001random}.

It should be noted that the $x$ and $y$ coordinates are predicted independently of one another. This requires training two separate random forests, $f_x$ and $f_y$. Both random forests use the same input data $v$, but are trained to predict different output coordinates. In other words, for an event $\mathbf{v}$, if $\hat{x} = f_x(\mathbf{v})$ and $\hat{y} = f_y(\mathbf{v})$, the final prediction resulting from the model will be $(\hat{x}, \hat{y})$.

\begin{figure*}[h]
\begin{center}
\includegraphics[width=0.75\linewidth]{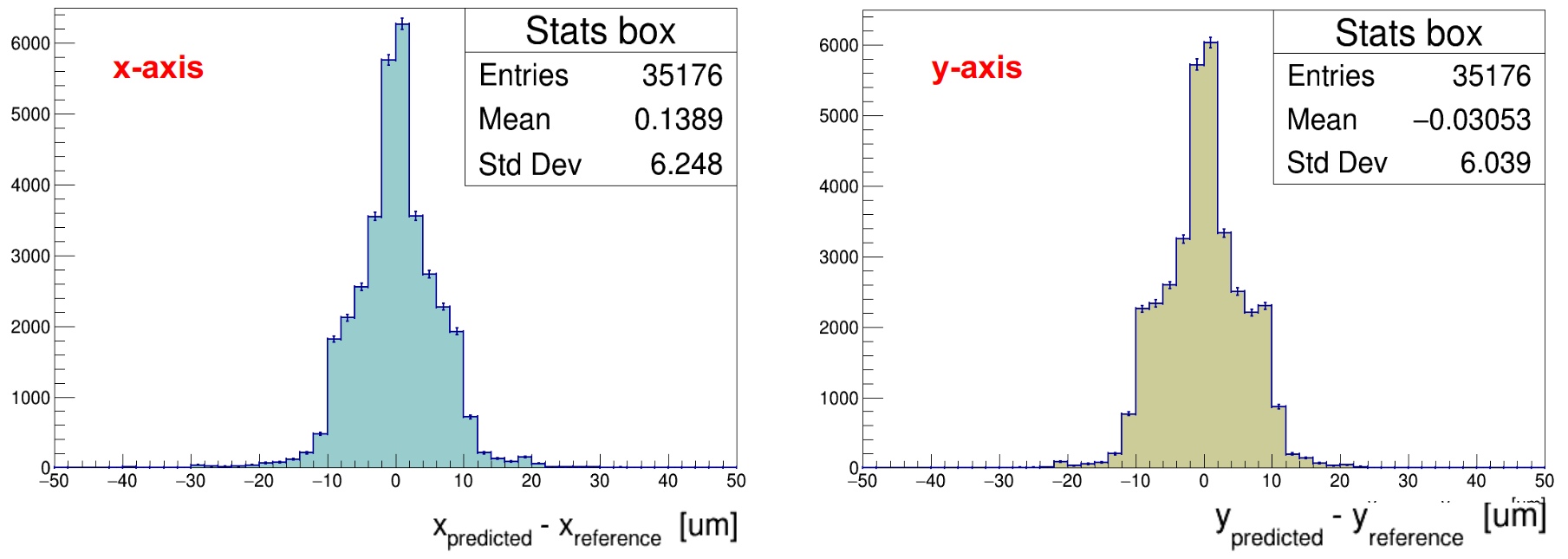}     
\caption{Distributions of the residuals, on both the $x$- and $y$-axis. $x_{predicted}$ ($y_{predicted}$) refers to the coordinate predicted by the ML algorithm on the $x$- ($y$-) axis; $x_{reference}$ ($y_{reference}$) is the laser reference position provided by the TCT setup.}
\label{fig:2_res}
\end{center}
\end{figure*}

\section{Experimental results}
\label{sec:results}
The spatial resolution has been computed on both the $x$- and $y$-axis, by comparing the positions predicted by the trained random forest to the laser reference positions provided by the $x-y$ stage. In particular, the standard deviation ($\sigma$) of the residuals distribution ($x_{predicted} - x_{reference}$ or $y_{predicted} - y_{reference}$) is a measurement of the spatial resolution of the entire system:

\begin{equation}
    \sigma_{total} = \sqrt{\sigma_{RSD}^{2}+\sigma_{laser}^{2}}
\end{equation}

from which it is possible to derive $\sigma_{RSD}$, knowing that $\sigma_{laser}$~\simile~2~\mum.

The results are very similar on both axes: $\sigma_{RSD}$~\simile~5.5~\mum, meaning that 91\% of the reconstructed positions are within $\pm$10~\mum~from the laser reference position. Those results are obtained from about 3.5~$\cdot$~10$^{4}$ events.

\begin{figure}[h]
\begin{center}
\includegraphics[width=9cm, height=4.5cm]{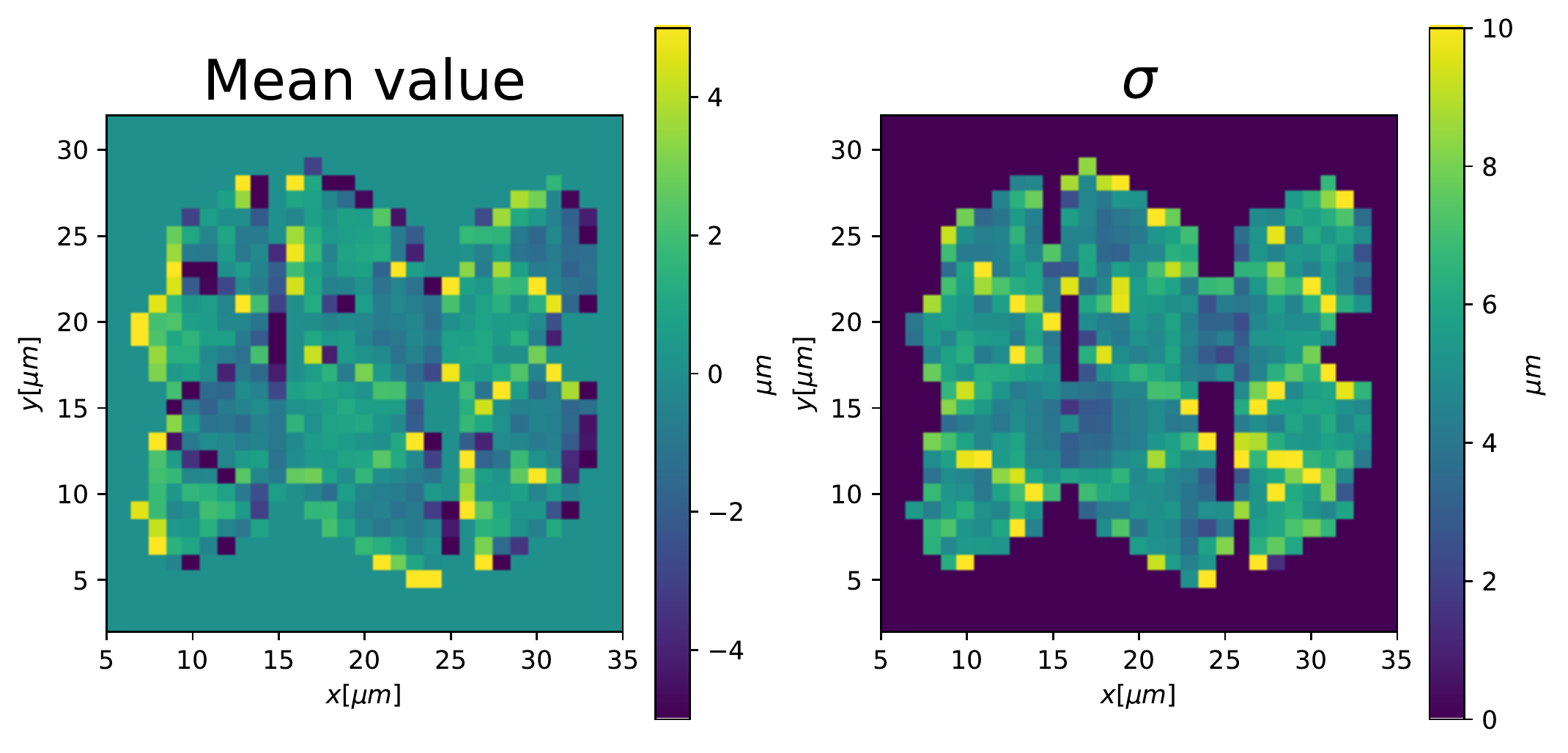}     
\caption{2d-map showing the mean value ($left$) and $\sigma$ ($right$) of the residuals distribution on the $x$-axis. Each pixel represents a $20 \times 20~\mu m^2$ bin.}
\label{fig:2d_maps}
\end{center}
\end{figure}

As illustrated in figure~\ref{fig:2_res}, the distributions of the residuals, on both axes, are not gaussian. We investigated that by dividing the DUT active area in 20~$\times$~20~\mum$^{2}$ bins and calculating the mean and $\sigma$ of the residuals in each bin. The results are shown in figure~\ref{fig:2d_maps} for what concerns the $x$-axis: despite the overall good uniformity, some differences can be noted in either the mean value or the $\sigma$.

\begin{figure}[h]
\begin{center}
\includegraphics[width=0.7\linewidth]{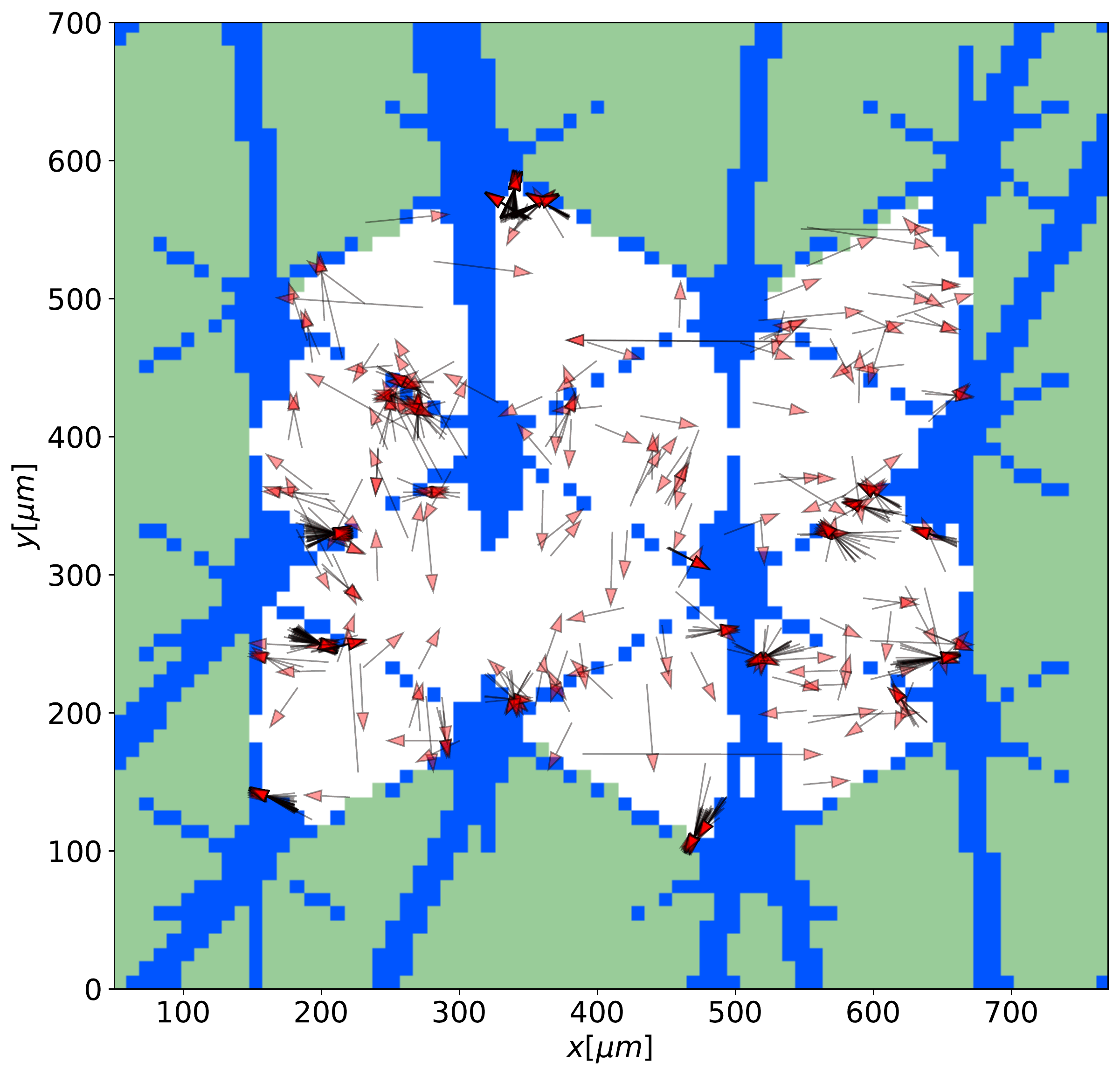}     
\caption{Map of the wrong predictions (distance predicted - reference positions $>$~20~\mum) on the test set. In blue are the wire bonds and the metal pads, in white the area of the sensor considered during the experiments, in green the portion of the sensor not under analysis.}
\label{fig:arrows}
\end{center}
\end{figure}

Figure~\ref{fig:arrows} offers an additional visualization of the distribution of the residuals. The arrows represent the events for which the predicted coordinates (tail of each arrow) have a distance larger than 20~\mum~from the reference coordinates (head of each arrow). It can be qualitatively seen that there are small local clusters of points that cause the majority of wrong predictions. 

Both figure~\ref{fig:arrows} and \ref{fig:2d_maps} illustrate that the regions where the positions are reconstructed worse are close to the arms of the metal pads, because the laser is partially absorbed there. In such regions, the distribution of the residuals have large $\sigma$ and/or mean values different from zero, causing the final distribution to be non-gaussian. 

The RSD spatial resolution is thus reported as the standard deviation of the distributions shown in figure~\ref{fig:2_res}; no results from the fit are presently considered.

\section{Conclusions}
\label{sec:conclusions}

We presented a 55~\mum-thick sensor coming from the FBK RSD2 production, with an active area of 750~$\times$~750~\mum$^{2}$ and 200~\mum~pitch, capable of achieving an excellent spatial resolution of less than 6~\mum, as measured with a precise TCT laser setup. As a comparison, a standard silicon pixel detector with the same pitch size and binary read-out would have a resolution of 30-60~\mum~\footnote{The spatial resolution of silicon pixel detectors with binary read-out is usually quoted as $k\cdot pitch/\sqrt{12}$, with $k$ = 0.5-1.}, a factor 5-10 higher than the RSD. The RSDs are thus very promising detectors for future trackers.

The key ingredients to achieve such an excellent position resolution are the built-in signal sharing and the internal charge multiplication, which allow large signals to be seen on multiple read-out pads.

The nature of the RSD signal sharing calls for the use of machine learning techniques, which are known to give better results than more standard techniques based on analytical sharing laws. In particular, a Random Forest gave the best results so far.

In the near future, more sophisticated techniques, such as fully-connected neural networks, will be considered, as well as new designs and wider sets of input features, in order to push further the performance of these innovative detectors.

\section{Acknowledgements}
We kindly acknowledge the following funding agencies and collaborations: INFN–CSN5, RSD Project;  FBK-INFN collaboration framework; MUR PRIN project 4DInSiDe; Dipartimenti di Eccellenza, Torino University (ex L.232/2016, art. 1, cc. 314, 337); SmartData@PoliTo.


\begin{thebibliography}{10}
\expandafter\ifx\csname url\endcsname\relax
  \def\url#1{\texttt{#1}}\fi
\expandafter\ifx\csname urlprefix\endcsname\relax\def\urlprefix{URL }\fi
\expandafter\ifx\csname href\endcsname\relax
  \def\href#1#2{#2} \def\path#1{#1}\fi
  
\bibitem{MM2} M. Mandurrino et al., \emph{Analysis and numerical design of Resistive AC-Coupled Silicon Detectors (RSD)}, \emph{NIM A} {Vol. A959} (2020).

\bibitem{marta} M.Tornago et al., \emph{{Resistive AC-Coupled Silicon Detectors: principles of operation and first results from a combined analysis of beam test and laser data}}, NIM A, \textbf{1003} (2021), 165319.

\bibitem{MM_RSD1} M.Mandurrino et al., \emph{{Demonstration of 200, 100, and 50 ~\mum pitch Resistive AC-Coupled Silicon Detectors (RSD) with 100\% fill-factor for 4D particle tracking}},  
IEEE Electron Device Letters, \textbf{40} (2019), no. 11.

\bibitem{MM_RSD2} M.Mandurrino et al., \emph{{The second production of RSD (AC-LGAD) at FBK}},  
\newblock \href {http://arxiv.org/abs/2111.14235}
  {\path{arxiv.org/abs/2111.14235}}.
  
\bibitem{fede_ml} F.Siviero et al., \emph{{First application of machine learning algorithms to the position reconstruction in Resistive Silicon Detectors}}, 2021 JINST \textbf{16} P03019.

\bibitem{TCT} \href{http://particulars.si} {\path{http://particulars.si}}

\bibitem{breiman2001random} L. Breiman, \emph{{Random forests}}, Machine Learning, \textbf{45} (2001), no. 1.

\bibitem{cern-seminar} N.Cartiglia, M.Mandurrino, \emph{Innovative Silicon Sensors for Future Trackers}, {CERN Detector Seminar} (2020).

\end{thebibliography}
\end{document}